\documentclass[]{aa}
\usepackage{graphicx}
\usepackage{txfonts}
\begin{document}

\title{Detection of lithium in nearby young late-M dwarfs}

   \author{N. Phan-Bao
       \inst{1,2}
       \and
       M.S. Bessell
       \inst{3}
       \and
       D. Nguyen-Thanh
       \inst{1,4}
       \and
       E.L. Mart\'{\i}n
       \inst{5}
       \and
       P.T.P. Ho
       \inst{2}
       \and
       C.F. Lee
       \inst{2}
       \and
       H. Parsons
       \inst{6}
       }

\offprints{N.~Phan-Bao}

\institute{Department of Physics, HCM International University-Vietnam National University, Block 6, 
           Linh Trung Ward, Thu Duc
           District, HCM city, Viet Nam. \\
           \email{pbngoc@hcmiu.edu.vn}
          \and
          Institute of Astronomy and Astrophysics, Academia Sinica. 
          PO Box 23-141, Taipei 106, Taiwan, ROC. 
          \and
          Research School of Astronomy and Astrophysics, Australian National University, 
          Cotter Rd, Weston, ACT 2611, Australia.
          \and         
          Faculty of Physics and Engineering Physics,
          HCM University of Science-VNU, 227 Nguyen Van Cu Street, 
          District 5, HCM city, Viet Nam.
          \and
          Centro de Astrobiologia (CSIC-INTA), Ctra. Ajalvir km 4, 28850 Torrej\'on de Ardoz, 
          Madrid, Spain.
          \and
          East Asian Observatory,
          660 N. A'Ohoku Place
          University Park,
          Hilo, Hawaii 96720, USA.  
          }

      \date{Received; accepted}

\abstract
{
Late M-type dwarfs in the solar neighborhood include a mixture of 
very low-mass stars and brown dwarfs which is 
difficult to disentangle due to the lack of constraints on their
age such as trigonometric parallax, lithium detection and space velocity. 
}
{We search for young brown dwarf candidates among a sample of 28 nearby late-M dwarfs
with spectral types between M5.0 and M9.0, 
and we also search for debris disks around three of them.
}
{Based on theoretical models, we used the color $I-J$, the $J$-band absolute magnitude and
the detection of the Li~I 6708~\AA~ doublet line as a strong constraint to estimate 
masses and ages of our targets.   
For the search of debris disks, we observed three targets at submillimeter wavelength
of 850~$\mu$m. 
}
{We report here the first clear detections of lithium absorption
in four targets and a marginal detection in one target.
Our mass estimates indicate that two of them are young brown dwarfs,
two are young brown dwarf candidates and one is a young very low-mass star.
The closest young field brown dwarf in our sample 
at only $\sim$15~pc is an excellent benchmark
for further studying physical properties of brown dwarfs in the range 100$-$150~Myr. 
We did not detect any debris disks around three
late-M dwarfs, and we estimated upper limits to the dust mass of debris disks around
them.
}
{}

\authorrunning{Phan-Bao et al.}
\titlerunning{Lithium in nearby, young late-M dwarfs}

  \maketitle

\section{Introduction}

Since the discovery of the first lithium-bearing late-M substellar-mass members 
in the benchmark Pleiades cluster (\citealt{rebolo95, rebolo96, basri96, martin96}), 
hundreds of brown dwarfs (BD, 13--75~$M_{\rm J}$) and very low-mass 
(VLM, 0.075--0.35~$M_{\odot}$) stars have been identified in the field 
and in young open clusters and star-forming regions.
 
According to the theory of VLM objects, a BD with a mass below 
$\sim$60~$M_{\rm J}$ should never reach a high enough core temperature to
destroy its primordial lithium content \citep{magazzu93}. 
Stars with masses below the mass limit of $\sim$0.35~$M_{\odot}$
(spectral types of $\sim$M3--M4) are predicted to be fully convective \citep{chab}.
Because BDs have masses well below this mass limit,
therefore, if lithium is not destroyed in the BD core, 
it should be detected in 
their atmosphere as initially proposed in \citet{rebolo92}.
However, the lithium depletion depends not only on stellar mass but
also on age, metallicity \citep{chab96} and rotation (e.g., \citealt{messina}). 
Some G, K and early-M dwarfs at young ages also exhibit
strong lithium absorption \citep{bopp,favata,song,murphy}. 
The combination of mass with age and lithium depletion
will set up a temperature boundary below which an object must
be substellar if lithium is detected. 
\citet{basri00} has shown that if lithium is present in any objects 
with effective temperatures below 2790~K, which corresponds to 
a spectral type of $\sim$M6 (e.g., \citealt{raj}),
these objects should be BDs (see Figure~1 in the Basri paper). 
Therefore, late-M dwarfs with spectral types of M6 or later
that show the Li~I resonance doublet line at 6708 \AA~ in their spectrum should be BDs.

Lithium as an age indicator has been used to search for young BDs among nearby late-M dwarfs. 
\citet{martin94} looked for lithium in 12 field late-M dwarfs without success.  
The first 3 lithium detections  
in field late-M dwarfs were reported in \citet{thack, tinney98, martin99a}. 
\citet{reid02} searched for lithium in 39 dwarfs with spectral type in the range M6.5--L0.5 and 
found strong lithium absorption in two late-M dwarfs. 
\citet{reiners09} searched for lithium in a sample of 63 late-M dwarfs and 
reported detections in six of them.  
Those searches for lithium among field late-M dwarfs indicate 
that about 10\% of them are young brown dwarfs.  

In this paper, we present a search for lithium in a sample of 28 nearby late-M dwarfs 
(spectral types from M5.0 to M9.0) selected on the basis of relatively 
strong H$\alpha$ emission in 
low-resolution spectra. The Li~I resonance doublet line centered at 6708 \AA~ is 
clearly detected in 
four and marginally detected in one
of our targets. We then used theoretical models to estimate the mass and the age of these 
lithium BD candidates.  These nearby BD candidates are benchmarks for further 
studies of the basic properties of young substellar mass objects.  

The rest of this paper is organized as follows: 
We present our selection of targets in Sect.~2 and the spectroscopic observations
in Sect.~3. In Sect.~4, we estimate spectral types and spectroscopic distances of
ten late-M dwarfs within 13~pc and then present the lithium detections and equivalent width 
measurement of H$\alpha$ and the Na~I doublet at 8183~\AA~ and 8199~\AA~ in our targets. 
We also discuss the variability of H$\alpha$ emission and report a strong flare observed in an M7.5 dwarf.
In Sect.~5, we estimate the masses and the ages of the five dwarfs with detected lithium
and present our search for debris disks around three targets. 
Section~6 summarizes our results. 

\section{Targets}
Because the presence of lithium in late-M dwarfs indicates 
that the sources are young and
their masses are substellar or very close to the substellar boundary,
we therefore selected late-M dwarfs with spectral types of $\ge$M5.0 (see Table~\ref{tab_EW}).
These late-M dwarfs were identified from the DENIS survey 
(see \citealt{pb01,pb03,pb06,crifo})
and have estimated spectral types 
as well as spectroscopic distances \citep{pb06,crifo}.
Most of the selected targets show relatively strong 
H$\alpha$ emission, 
an indicator of magnetic activity. 

Young late-M dwarfs are
magnetically active and thus show strong H$\alpha$ emission.
However, in late-M dwarfs, the magnetic activity
depends not only on their age but also on the dynamo mechanism operating 
in these fully convective stars (e.g., see \citealt{pb09} and references therein). 
In addition,  
\citet{west08} have suggested that the lifetimes of magnetic activity 
of late-M dwarfs may be a few Gyr. Many old late-M dwarfs also show strong 
H$\alpha$ emission.
Therefore, one should note that the presence of H$\alpha$ emission in our late-M dwarfs
can not confirm their youth but it implies that the selected targets are 
potential young late-M dwarf candidates.
 
In our sample, we also included very nearby late-M dwarfs that we had identified 
in the DENIS database (see \citealt{pb08} and references therein)
with photometric distances within 12~pc based on the $M_{\rm I}$ absolute magnitude
versus $I-J$ color relationship in \cite{pb03}.
These dwarfs have been identified
in high-proper motion surveys (see Table~\ref{tab_SpT}), 
although several lack spectroscopic/trigonometric distances
(SCR J0838$-$5855, PM J14223$-$7023, SCR J1546$-$5534,
PM J17189$-$4131, SIPS J1809$-$7613, SCR J1855$-$6914 and LEHPM 5031)
or spectroscopic spectral types (SCR J0838$-$5855, PM J17189$-$4131, 
SIPS J1809$-$7613, SCR J1855$-$6914 and LEHPM 5031).
\section{Spectroscopic observations}

We observed twenty eight late-M dwarfs from 2005 to 2008 with the double-beam grating
spectrograph (DBS) on
the 2.3 m telescope at Siding Spring Observatory. The red channel of the DBS covers the
wavelength range of 6480-7485 \AA. 
The 1200~g/mm grating was used, providing a
medium-resolution of about 1.0~\AA, at 0.5~\AA/pixel.  
Table~\ref{tab_EW} lists the observing date of our targets. 

For 10 very nearby late-M dwarfs as listed in Table~\ref{tab_SpT} 
(except LHS~234 and SCR~J0838$-$5855), we observed them
with DBS with the 316~g/mm grating, which covers the wavelength range of 
6200$-$10050~\AA~at a low-resolution of about 4~\AA, at 2~\AA/pixel. 

We used FIGARO \citep{shortridge} to reduce the data. 
The data was flat fielded. No dark subtraction was done due to the 
insignificant dark current. 
No bad pixels were removed. 
One bad column was removed by interpolation.  
The 2D long slit spectrum of a bright star was traced and 
the image was transformed to straighten the spectrum. 
This transformation was then applied to each program star. 
All observations were made with the atmospheric 
dispersion along the slit so there is no curvature caused 
by the atmospheric dispersion.
Extremely metal-deficient red giants with temperatures
of 5000$-$6000~K and with [Fe/H]$<$$-$2 that make good smooth spectrum templates 
were observed during the night
at air masses encompassing those for the targets
to remove the telluric lines. 
These were then divided into the program star's spectrum
and the residuals were examined. 
The best divided result was accepted.
Each night one or two white dwarfs, such as EG~131, VMa~2, 
L745-46A, LTT4364 were additionally observed to test whether the telluric removal was well carried out
as they have no bands or lines redward of
4100~\AA~ and they are smooth blackbodies.
After flat fielding and telluric absorption line removal,  
the standard star spectra in the red are very smooth 
and show slowly varying changes of continuum intensity with wavelength.   
These standards were also used 
for flux calibration
and a NeAr arc for wavelength calibration.
The details of the technique were given in \citet{bessell99}. 

The signal-to-noise ratios are in the range of 7$-$23 for medium-resolution spectra,
except four spectra (J0041353$-$562112, J0103119$-$535143, J0517377$-$334903
and J1357149$-$143852) with moderate values of 3$-$5.
For low-resolution spectra, the signal-to-noise ranges from 8 to 20. 
\begin{figure*}
   \centering
    \includegraphics[width=15cm,angle=-90]{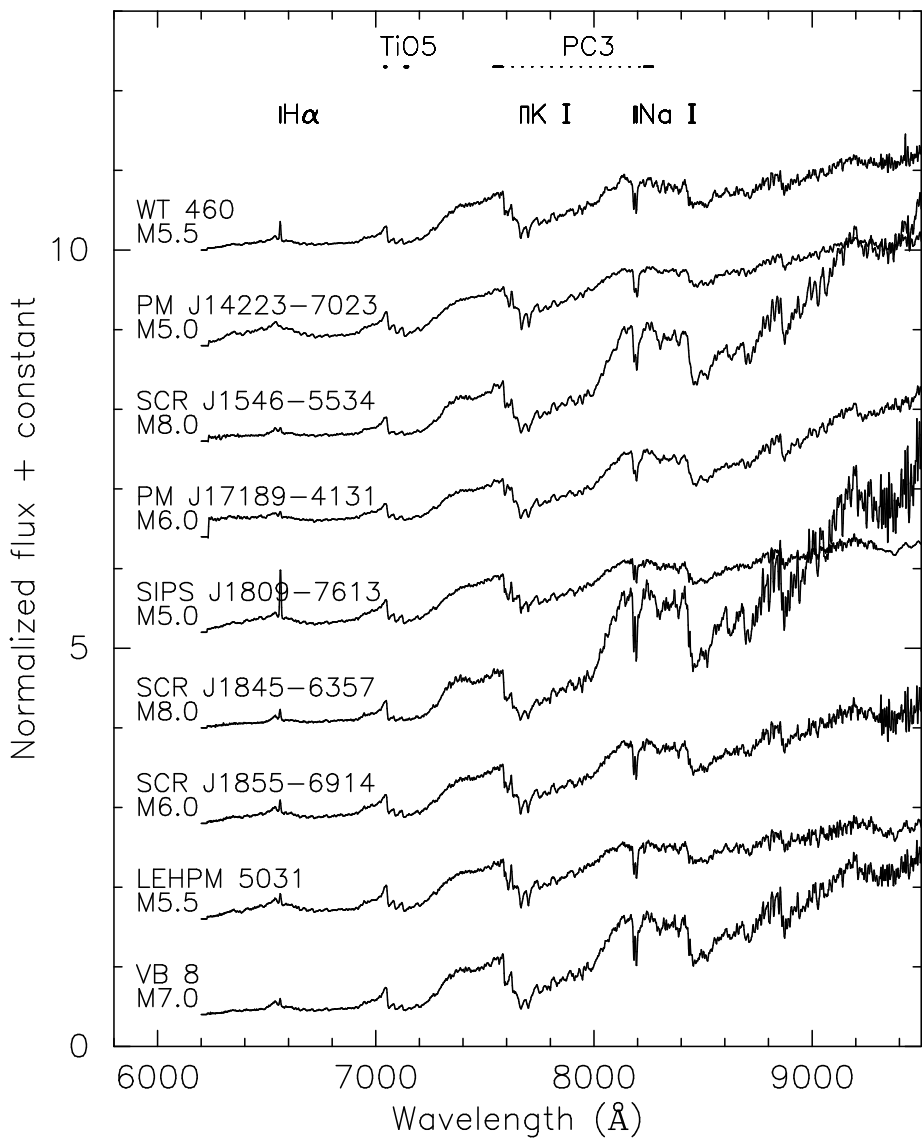}
    \caption{Low-resolution spectra of 8 very nearby late-M dwarfs and the VB~8 reference 
as listed in Table~\ref{tab_SpT}. The H$\alpha$, Na~I, K~I lines and the spectral intervals used
for calculating the TiO5 and PC3 indices 
are indicated. }
\label{f1}
\end{figure*}
\begin{figure*}
   \centering
    \includegraphics[width=15cm,angle=-90]{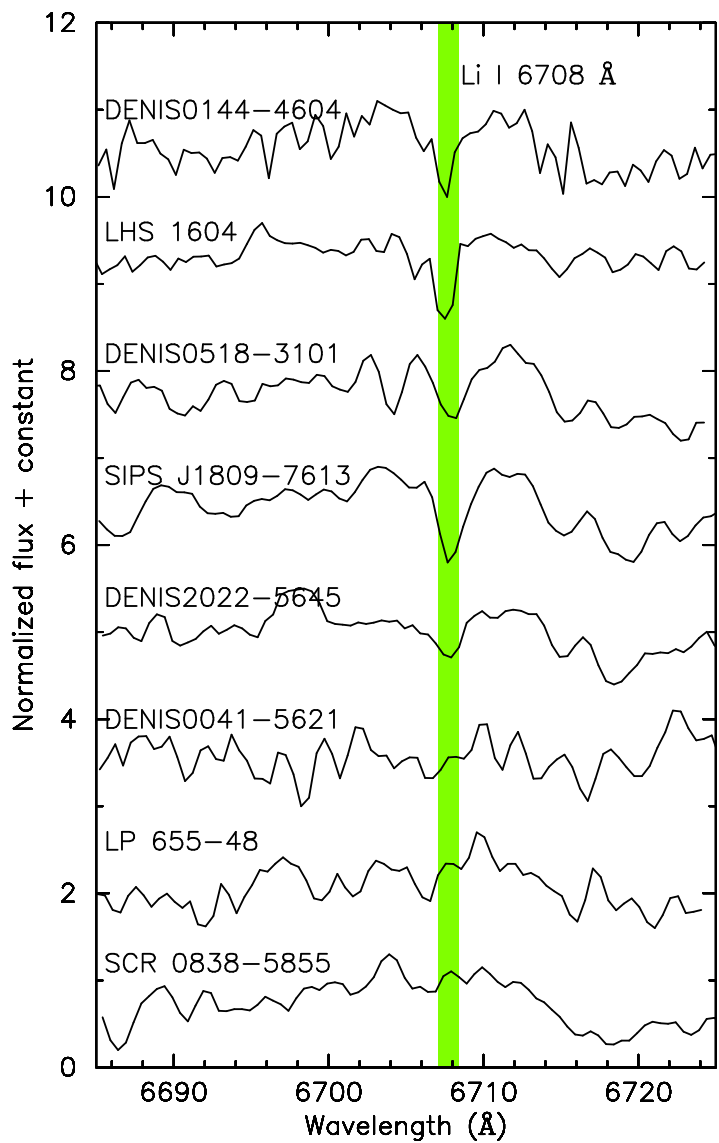}
    \caption{Medium-resolution spectra of five late-M dwarfs with detected lithium
(five upper spectra). The spectra
of three late-M dwarfs with non-detection of lithium are also shown (three lower spectra). 
The region of the Li~I resonance doublet
line at 6708~\AA~ \citep{pav} is indicated.
}
\label{f2}
\end{figure*}
\section{Results}
\subsection{Estimate of spectral types and distances of 10 very nearby late-M dwarfs}
Based on both low- and medium-resolution spectra,
we estimated the spectral types of the 10 very nearby late-M dwarfs (Table~\ref{tab_SpT}).
Spectral types of M dwarfs could be determined using
spectral indices PC3 \citep{martin99b}, 
TiO5 and VOa (see \citealt{cruz02} and references therein), 
as described in detail in \citet{pb06}. 
The adopted spectral type is an average value of three spectral types
estimated from the three spectral indices, except some cases as mentioned
below.
The PC3 and VOa indices were measured using low-resolution spectra.
The TiO5 index is sensitive to 
the spectral resolution because the wavelength interval for the TiO5 denominator is very
narrow (only 4~\AA, 7042$-$7046~\AA) and positioned at 
the head of a molecular band as explained in \citet{crifo}.
We therefore used our medium-resolution spectra available for 7 targets 
(Table~\ref{tab_SpT}) to measure TiO5.  
For the 3 remaining targets, low-resolution spectra were used to determine TiO5.
For the cases of LHS 234 and
SCR J0838$-$5855, 
because we did not observe these two targets at low-resolution spectroscopy,
therefore, only the TiO5 index was used to estimate their spectral type.
The PC3 and VOa indices are not available for
LHS 234 and SCR J0838$-$5855 as 
the observed wavelength of medium-resolution spectroscopy does not cover these indices.

To compute the distances, we used the PC3 index versus absolute magnitude 
relations in $I$, $J$ and $K$ bands \citep{crifo}, 
or the spectral type versus $J$-band absolute magnitude relation
\citep{filip} for LHS 234 and SCR J0838$-$5855. 
Table~\ref{tab_SpT} lists spectral types and distances
and their associated errors estimated
for these 10 late-M dwarfs.   
One should note that these relations are applicable for
field age late-M dwarfs. 
If our objects are young ($\sim$10$-$150 Myr), 
our absolute magnitudes 
may be overestimated by 0.5$-$2.0 mags \citep{faherty,filip}, which results
in the distances being underestimated by 23$-$92\%.

For the case of SCR J1546$-$5534 (or DENIS-P J1546418$-$553446, M8.0), 
our spectroscopic distance is 
only 5.4$\pm$0.7~pc, which is closer than the photometric distance of
6.7~pc estimated by \citet{boyd}.

\subsection{Lithium detection in five late-M dwarfs}
Based on medium-resolution spectra of 28 late-M dwarfs,
we clearly detected the Li~I resonance doublet line at 6708~\AA~ 
in four objects: DENIS-P J0144318$-$460432 (hereafter DENIS0144$-$4604), 
LHS~1604 (or DENIS-P J0351000$-$005244),  
SIPS J1809$-$7613 (or DENIS-P J1809068$-$761324) and 
DENIS-P J2022480$-$564556 (hereafter DENIS2022$-$5645),
as shown in Figure~\ref{f2}. 
For the case of DENIS-P J0518113$-$310153 (hereafter DENIS0518$-$3101),
the lithium is marginally detected.
Using the IRAF task splot, we manually measured 
equivalent widths (EW) of the Li~I line at 6708~\AA.
The continuum levels and integration limits were examined
individually for each spectrum. The uncertainties in the EW measurement
were derived by measuring EWs with different possible continuum levels
as well as examining the noise around the region of 6700-6720~\AA.
For the cases of non-detection of lithium, we measured upper limits
by examining the noise in the region of interest.
Our measurements are listed in Table~\ref{tab_EW}. 

Using the same manner as applied for the Li~I EW measurement, 
we measured H$\alpha$ EWs and their associated uncertainties for all targets. 
Four of the five late-M dwarfs with detected
lithium have shown a significant variation in H$\alpha$ emission
based on spectra obtained in this paper (see Tables~\ref{tab_EW} and \ref{tab_SpT}),
published in \citet{crifo,pb06} and reported in the literature 
at different epochs (UT time) as follows:
\begin{itemize}
\item DENIS0144$-$4604 has H$\alpha$ emission with EW = $-$12.1~\AA~ measured in 2005-07-29 
and $-$25.4~\AA~ in 2005-07-30 \citep{pb06}. 
\item LHS~1604 has EW H$\alpha$ = $-$5.9~\AA~ measured in both 2006-01-10 and 2003-11-29 \citep{crifo}
but $-$11.8~\AA~ in 2003-10-18 \citep{west}.
\item DENIS0518$-$3101 has EW H$\alpha$ = $-$8.4~\AA~ measured in 2008-03-28 and
$-$18.8~\AA~ in 2003-11-29 \citep{crifo}.
\item SIPS J1809$-$7613 has EW H$\alpha$ =  $-$4.7~\AA~ measured in 2008-03-28 and
$-$23.2~\AA~ in 2007-08-04.
\end{itemize}
The variable H$\alpha$ emission in these late-M dwarfs
is possibly due to either flaring activity  
or their rotation (e.g., \citealt{berger08a,berger08b,pb09}). 
For the case of
DENIS2022$-$5645, its H$\alpha$ emission was likely stable,
with EW = $-$5.3~\AA~ measured in
2008-03-28 and $-$5.7~\AA~ in 2002-08-06 \citep{crifo}.
The stable H$\alpha$ emission suggests that this dwarf is (nearly) pole-on
and there was no flare during the observations.
Spectroscopic monitoring of these five dwarfs for a full rotational period 
will clarify the possibilities. 

\begin{figure}
   \centering
    \includegraphics[width=10.0cm,angle=-90]{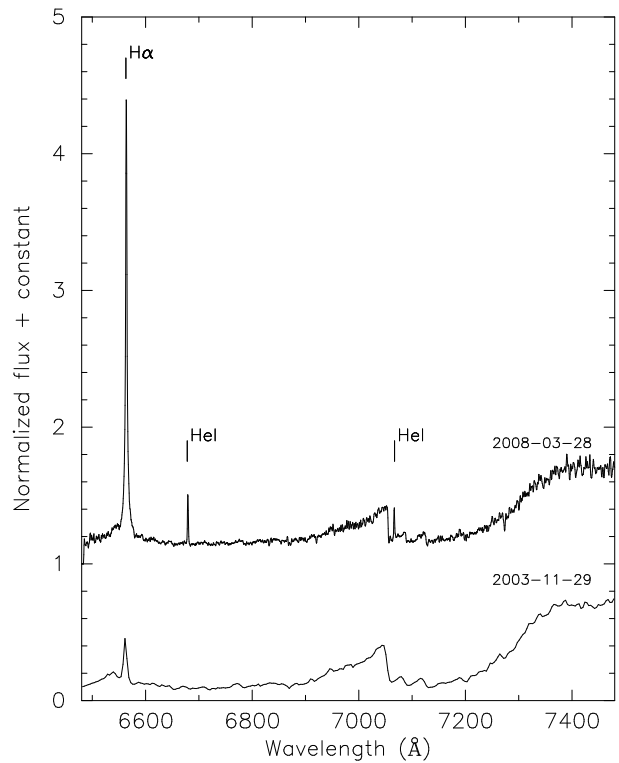}
    \caption{LP 655-48 (M7.5, 9.5~pc) in a strongly flaring level with very strong
H$\alpha$ and He~I emission lines (upper spectrum, this paper) 
and in a lower level of activity (lower spectrum, \citealt{crifo}).
The observing dates are also indicated.
}
\label{f3}
\end{figure}

In addition, we note a strong flare observed in LP~655-48 (M7.5) 
during our observations (see Figure~\ref{f3}) with
EW H$\alpha$ = $-$35.2~\AA~ measured in 2008-03-28.
This source has EW H$\alpha$ = $-$13.6~\AA~ measured in 2003-11-29
from the spectrum taken in \cite{crifo}.
This source also showed strong He~I emission lines  
as seen in LP~412-31 (M8, \citealt{schmidt}).
One should discuss here that we detected no lithium
in the source (Figure~\ref{f2}), which is in agreement with 
the non-detection of lithium as reported in
\citet{reiners09}. 
LP 655-48 has a trigonometric distance of 9.5$\pm$0.3~pc \citep{shkol12},
which is in agreement with its spectroscopic distance of 8.9$\pm$1.3~pc \citep{crifo}. 
With $J=10.74$, using the trigonometric distance of the source 
we then derived $M_{\rm J} = 10.85$. 
According to the CIFIST2011 BT-Settl atmosphere models \citep{allard13},
the $J$-band absolute magnitude and the DENIS color $I-J = 2.61$ of the source
imply its age $>$400 Myr. This age estimate is likely
consistent with the identification of LP 655-48 as a candidate
member of Hyades ($\sim$600$-$800~Myr, \citealt{brandt}) 
by \citet{go10}.
At an age $>$400 Myr, the source should have no lithium (see also Figure~\ref{f5}). 
This is consistent with the non-detections of lithium
but disagrees with the detection of weak lithium absorption  
as reported in \citet{shkol09} and an age range of 
10$-$90~Myr estimated in \citet{shkol12}.
One should note that
the source has a radial velocity of 31.1$\pm$1.4~km~s$^{-1}$ \citep{deshpande}. 
Using BANYAN II\footnote{http://www.astro.umontreal.ca/$\sim$gagne/banyanII.php}, 
which is a Bayesian analysis tool to estimate the membership probability of candidates 
to nearby young moving groups (see \citealt{gagne14,malo}),
and the proper motion of LP 655-48 in \citet{pb03}, 
we found that the object has a 75.11\% probability candidate member of old field (age $>$1~Gyr). 
Therefore, LP 655-48 is very likely an old field late-M dwarf.

Using the spectra of 28 late-M dwarfs published in \citet{crifo} and \citet{pb06},
we also measured EWs of the Na~I doublet at 8183~\AA~ and 8199~\AA,
which is an indicator for surface gravity of young M dwarfs (e.g., \citealt{lyo,schlieder}),
over the range of 8170$-$8200~\AA~ as described in \citet{martin10}.
The uncertainties in the EW measurement
were estimated by measuring EWs with different possible continuum levels.
Our measurements are listed in Table~\ref{tab_EW}.
Figure~\ref{f3} shows the Na~I EW
versus spectral type relation for our late-M dwarfs. 
Three of five dwarfs with detected lithium show weak Na~I: DENIS0144$-$4604,
SIPS J1809$-$7613 and DENIS2022$-$5645. 
Their Na~I EWs are comparable to those of 7 Upper Sco candidates 
measured in \citet{martin10}. 
This implies that they
have low surface gravity. 
The Na~I EW of DENIS0144$-$4604 (EW = 3.5~\AA, M5.5)
is comparable to that of DENIS0041$-$5621 (EW = 3.3~\AA, M7.5).
At this point, one should note that 
DENIS0041$-$5621, which was identified as an M7.5 \citep{pb01,pb06},
is actually a binary of M6.5+M9.0 \citep{reiners10}
with lithium detection as reported in 
\citet{reiners09} and \citet{bur15} (EW = 0.7~\AA).
However, we did not detect lithium in DENIS0041$-$5621 (see Figure\ref{f3})
with an upper limit of 0.2~\AA. 
The non-detection of lithium in the source
is probably due to its low spectral signal-to-noise
ratio of only $\sim$3. 
We therefore cannot confirm the previous detections of
lithium in DENIS0041$-$5621 with our current data.
The two remaining ones, LHS~1604 and DENIS0518$-$3101, show Na~I EWs significantly
higher than those of the 3 objects above. 
For the case of LHS~1604, no lithium has been detected
in the previous observations \citep{reiners09,bur15}. 
However, we detected a strong lithium absorption line with EW = 1.2$\pm$0.2~\AA~ (see Figure~\ref{f2})
which is right at the upper limit of 1.2~\AA~ as reported in \citet{bur15}. 

\begin{table*}
   \caption{H$\alpha$, Li~I 6708 \AA~ and Na~I (8170-8200~\AA) equivalent widths of 28 nearby late-M dwarfs.}
    \label{tab_EW}
  $$
   \begin{tabular}{lllllllll}
   \hline 
   \hline
   \noalign{\smallskip}
DENIS-P name               &  Other name   & SpT & Distance &  UT observing  &   {\it EW} H$\alpha$ & {\it EW} Li & {\it EW} Na~I &  References \\
                         &               &     & (pc)     &   date      & (\AA)          &  (\AA)        &  (\AA)                  & \\
      \noalign{\smallskip}
\hline
J0041353$-$562112$^{\rm a}$  &        &  M7.5 &  17.0$\pm$2.4 & 2008-03-29 &$-$16.1$\pm$0.4 &  $<$0.2       &  3.3$\pm$0.6 &  (1)    \\
J0103119$-$535143  &                  &  M5.5 &  24.3$\pm$3.7 & 2008-03-29 &~~$-$8.9$\pm$0.1 &  $<$0.1       &  5.0$\pm$0.1 &  (1)    \\
J0144318$-$460432  &                  &  M5.5 &  23.3$\pm$3.4 & 2005-07-29 &$-$12.1$\pm$0.2 &   0.4$\pm$0.1 &  3.5$\pm$0.6 &  (1)    \\
J0253444$-$795913  &                  &  M5.5 &  17.2$\pm$2.4 & 2005-07-29 &~~$-$6.5$\pm$0.1 &  $<$0.2       &  5.7$\pm$0.6 &  (1)    \\
J0334113$-$495333  &                  &  M9.0 & ~~8.2$\pm$0.8 & 2005-11-09 & >$-$0.1       &  $<$0.1       &  6.7$\pm$0.1 &  (2)    \\
J0351000$-$005244  & LHS 1604         &  M7.0 &  12.8$\pm$1.8 & 2006-01-10 &~~$-$5.9$\pm$0.4 &   1.2$\pm$0.2 &  7.6$\pm$0.5 &  (3)    \\
J0410480$-$125142  &LP 714-37$^{\rm b}$& M5.5 &  18.1$\pm$2.2 & 2008-03-28 &~~$-$0.3$\pm$0.1 &  $<$0.1       &  7.7$\pm$0.4 &  (4) \\
J0440231$-$053009  & LP 655-48        &  M7.5 & ~~8.9$\pm$1.3 & 2008-03-28 &$-$35.2$\pm$0.2 & $<$0.1        &  8.5$\pm$0.5 &  (3)  \\
J0517377$-$334903  &                  &  M8.0 &  12.1$\pm$1.8 & 2008-03-28 &~~$-$3.9$\pm$1.2 &  $<$0.3       &  5.9$\pm$0.6 &  (1)    \\
J0518113$-$310153  &                  &  M6.5 &  19.5$\pm$2.9 & 2008-03-28 &~~$-$8.4$\pm$0.3 &~~0.6$\pm$0.3$^{\rm e}$&  5.8$\pm$0.1 & (3)\\
J0740193$-$172445  & LHS 234          &  M6.5 & ~~9.1$\pm$1.3 & 2008-03-29 & >$-$0.1       &  $<$0.1       &              &  this paper \\
J0838022$-$585558  & SCR 0838$-$5855  &  M6.0 &  11.3$\pm$1.6 & 2008-03-29 &~~$-$3.2$\pm$0.1 &  $<$0.1     &              &  this paper  \\
J1236153$-$310646  & LP 909-55        &  M5.5 &  19.4$\pm$2.7 & 2008-03-28 &~~$-$8.3$\pm$0.2 &  $<$0.1       &  6.2$\pm$0.5 &  (1) \\
J1357149$-$143852  &                  &  M7.5 &  24.7$\pm$3.6 & 2008-03-28 &~~$-$5.9$\pm$0.9 &  $<$0.1       &  6.2$\pm$0.8 &  (1) \\
J1411599$-$413221  & WT 460$^{\rm c}$ &  M5.5 &  10.1$\pm$1.3 & 2008-03-28 &~~$-$6.2$\pm$0.1 &  $<$0.1       &  7.6$\pm$0.3 &  this paper\\
J1538317$-$103850  &                  &  M5.0 &  31.7$\pm$7.0 & 2008-03-28 &$-$13.1$\pm$0.1 &  $<$0.1       &  4.3$\pm$0.9 &  (3) \\
J1553571$-$231152  & LP 860-46        &  M5.0 &  21.5$\pm$2.9 & 2008-03-28 &~~$-$5.5$\pm$0.1 &  $<$0.2       &  7.2$\pm$0.5 &  (3) \\
J1610584$-$063132  & LP 684-33        &  M5.5 &  17.7$\pm$2.5 & 2008-03-28 &~~$-$3.5$\pm$0.1 &  $<$0.1       &  8.3$\pm$0.1 &  (3) \\
J1809068$-$761324  & SIPS J1809$-$7613&  M5.0 &  10.4$\pm$1.4 & 2008-03-28 &~~$-$4.7$\pm$0.1 &   0.5$\pm$0.1 &  4.4$\pm$0.2 & this paper\\
J1845049$-$635747  & SCR J1845$-$6357$^{\rm d}$& M8.0&~~3.2$\pm$0.4 & 2008-03-28 &~~$-$2.1$\pm$0.1 &  $<$0.1 &  9.0$\pm$0.5 & this paper\\
J1855480$-$691415  & SCR J1855$-$6914 &  M6.0 &  11.0$\pm$1.6 & 2008-03-28 &~~$-$3.5$\pm$0.1 &  $<$0.1       &  7.6$\pm$0.3 & this paper\\
J1917045$-$301920  & LP 924-17        &  M5.5 &  22.1$\pm$3.1 & 2008-03-28 &~~$-$8.5$\pm$0.2 &  $<$0.1       &  8.3$\pm$0.2 &  (3) \\
J2002134$-$542555  &                  &  M6.0 &  17.5$\pm$2.5 & 2005-07-29 &~~$-$3.4$\pm$0.1 &  $<$0.1       &  5.6$\pm$0.1 &  (1)  \\
J2022480$-$564556  &                  &  M5.5 &  22.9$\pm$3.3 & 2008-03-28 &~~$-$5.3$\pm$0.2 &   0.4$\pm$0.1 &  5.0$\pm$0.8 &  (3) \\
J2049527$-$171608  & LP 816-10        &  M6.0 &  19.4$\pm$5.7 & 2008-03-28 &~~$-$5.3$\pm$0.2 &  $<$0.1       &  8.6$\pm$0.6 &  (3) \\
J2132297$-$051158  & LP 698-2         &  M5.5 &  18.5$\pm$2.6 & 2005-07-29 & >$-$0.1       &  $<$0.2       &  5.5$\pm$0.2 &  (1)  \\
J2151270$-$012713  & LP 638-50        &  M5.0 &  18.7$\pm$3.1 & 2005-07-29 &~~$-$2.8$\pm$0.1 &  $<$0.1       &  5.0$\pm$0.1 &  (1)  \\
J2241593$-$750034  & LEHPM 5031       &  M5.5 &  12.5$\pm$1.8 & 2008-03-28 &~~$-$1.9$\pm$0.1 &  $<$0.1       &  8.5$\pm$0.8 & this paper\\
    \noalign{\smallskip}
    \hline 
   \end{tabular}
  $$
  \begin{list}{}{}
  \item[] 
{\bf Notes.} 
The H$\alpha$ and Li~I EWs measured from the medium-resolution spectra, and the NaI EWs 
measured from low-resolution spectra published in \citet{crifo} and \citet{pb06} (see Sect. 4.2).\\
$^{\rm (a)}$: A young binary of M6.5+M9.0 \citep{reiners10};
$^{\rm (b)}$: A triple system of M5.5+M8.0+M8.5 \citep{pb05,pb06b}; 
$^{\rm (c)}$: A binary with photometric spectral types of M6.0+L1 estimated in \citet{mon};
$^{\rm (d)}$: A binary of M8.5+T5.5 \citep{henry,bill}; $^{\rm (e)}$: the lithium is marginally detected.\\
References for spectral type and distance: (1) \citet{pb06}; (2) \citet{pb06a}; (3) \citet{crifo};
(4) \citet{pb05}.
  \end{list}
\end{table*}
\begin{table*}
   \caption{Spectral indices, distances and spectral types estimated for 10 very nearby late-M dwarfs 
and the VB 8 reference observed at MSSSO-2.3 m.}
    \label{tab_SpT}
  $$
   \begin{tabular}{lllllllllll}
   \hline 
   \hline
   \noalign{\smallskip}
Name              & $I$   &  $J$  & $K$   & UT observing & VOa  & TiO5 & PC3 & Distance & SpT$^{\rm c}$ & Ref. \\
                  &       &       &       &   date    &      &      &     &  (pc)    &     & \\ 
      \noalign{\smallskip}
\hline
LHS 234           & 12.37 & 10.21 &  9.31 & 2008-03-29 &      & 0.15$\pm$0.03$^{\rm a}$ &      & ~~9.1$\pm$1.3$^{\rm b}$ & M6.5 & 1 \\
SCR J0838$-$5855  & 12.77 & 10.34 &  9.20 & 2008-03-29 &      & 0.20$\pm$0.02$^{\rm a}$ &      & 11.3$\pm$1.6$^{\rm b}$ & M6.0 & 2 \\
WT 460            & 11.77 &~~9.64 &  8.59 & 2007-08-04 & 2.06$\pm$0.14 & 0.20$\pm$0.02$^{\rm a}$ & 1.22$\pm$0.09 & 10.1$\pm$1.3 & M5.5 & 3 \\
PM J14223$-$7023  & 11.97 & 10.15 &  9.43 & 2008-03-27 & 2.02$\pm$0.07 & 0.36$\pm$0.08 & 1.38$\pm$0.05 & 10.0$\pm$1.4 & M5.0 & 4 \\
SCR J1546$-$5534  & 12.96 & 10.22 &  9.14 & 2008-03-27 & 2.22$\pm$0.15 & 0.29$\pm$0.03 & 2.05$\pm$0.10 &~~5.4$\pm$0.7 & M8.0 & 5 \\
PM J17189$-$4131  & 12.95 & 10.55 &  9.60 & 2008-03-27 & 2.10$\pm$0.09 & 0.52$^{\rm d}$          & 1.52$\pm$0.05 & 10.5$\pm$1.5 & M6.0 & 4 \\
SIPS J1809$-$7613 & 11.69 &~~9.87 &  8.96 & 2007-08-04 & 2.02$\pm$0.12 & 0.26$\pm$0.03$^{\rm a}$ & 1.25$\pm$0.06 & 10.4$\pm$1.4 & M5.0 & 6 \\
SCR J1845$-$6357  & 12.51 &~~9.52 &  8.48 & 2007-08-04 & 2.21$\pm$0.20 & 0.28$\pm$0.03$^{\rm a}$ & 2.42$\pm$0.18 &~~3.2$\pm$0.4 & M8.0 & 7 \\
SCR J1855$-$6914  & 12.67 & 10.46 &  9.50 & 2007-08-04 & 2.10$\pm$0.16 & 0.22$\pm$0.02$^{\rm a}$ & 1.43$\pm$0.09 & 11.0$\pm$1.6 & M6.0 & 8 \\
LEHPM 5031        & 12.54 & 10.39 &  9.54 & 2007-08-04 & 2.04$\pm$0.11 & 0.24$\pm$0.03$^{\rm a}$ & 1.32$\pm$0.11 & 12.5$\pm$1.8 & M5.5 & 9 \\
VB 8              & 12.24 &~~9.74 &  8.82 & 2007-08-04 & 2.15$\pm$0.18 & 0.18$\pm$0.01 & 1.74$\pm$0.11 &~~5.8$\pm$0.8 & M7.0 &   \\
    \noalign{\smallskip}
    \hline 
   \end{tabular}
  $$
  \begin{list}{}{}
  \item[] 
{\bf Notes.} The VOa, TiO5 and PC3 indices measured from the low-resolution spectra, 
except TiO5 for some cases as noted (see Sect. 4.1). \\
$^{\rm (a)}$ The TiO5 index measured from medium-resolution spectra. \\
$^{\rm (b)}$ Distances estimated from the spectral type versus magnitude
relation in \citet{filip}. \\
$^{\rm (c)}$ An uncertainty of 0.5 subtypes of estimated spectral types. \\ 
$^{\rm (d)}$ An unreliable value, not being used to estimate spectral type. \\
References for the source name: (1) \citet{luyten}; (2) \citet{fin}; (3) \citet{wro};
(4) \citet{lepine}; (5) \citet{boyd}; (6) \citet{deacon}; (7) \citet{hambly}; 
(8) \citet{suba}; (9) \citet{pok}.
  \end{list}
\end{table*}
\section{Discussion}
The detection of the Li~I 6708~\AA~ doublet line in five late-M dwarfs
DENIS0144$-$4604 (M5.5), LHS~1604 (M7.0),  DENIS0518$-$3101 (M6.5), 
SIPS1809$-$7613 (M5.0) and DENIS2022$-$5645 (M5.5)
indicates that these dwarfs are possibly young BDs.
In order to determine the substellar nature of these objects, 
we used the CIFIST2011 BT-Settl atmosphere models \citep{allard13} 
for the DENIS photometric system
to estimate their mass 
and age range.

\subsection{Estimate of mass and age range of five young VLM objects}
{\it DENIS0144$-$4604 (M5.5):} 
This late-M dwarf was discovered by \cite{pb03} and 
spectroscopically classified as an M5.5 in \citet{pb06}. 
No trigonometric parallax and radial velocity measurements 
have been reported. 
The detection of lithium in the object will place the source
in the lithium region (see Fig.~\ref{f5}). This thus
indicates that 
its real absolute magnitude should be brighter 
than the magnitude derived from its spectroscopic distance 
(see Table~\ref{tab_EW}). 
According to the BT-Settl models,
the lithium presence in DENIS0144$-$4604
and its color $I-J = 2.19$ \citep{pb03}
indicate its age $\le$120 Myr. 
This age constraint places an upper limit of $\sim$73~$M_{\rm J}$ 
to the mass of DENIS0144$-$4604. 
The source is therefore a young BD. 
Using the BANYAN II tool and the proper motion measured in
\citet{pb03}, we found that DENIS0144$-$4604 has a 97.07\% probability candidate 
member of the Tucana-Horologium moving group and a kinematic distance of 40$\pm$3~pc. 
Assuming the source at this distance, with an apparent
$J$-band magnitude of 11.91 \citep{pb03} we then derived
its absolute magnitude $M_{\rm J} = 8.90$.
Based on the BT-Settl models, 
with such an absolute magnitude
the object should have an age range of 10$-$20~Myr (Fig.\ref{f5}).
This age range however is not consistent with the Tucana-Horologium
age of 45$\pm$4~Myr \citep{bell}. 
Therefore, measurements of radial velocity and/or parallax of 
DENIS0144$-$4604 are clearly needed to determine its mass precisely and
its membership in nearby young moving groups.

{\it LHS 1604 (M7.0):} The source has a spectral type of M7.0 
(e.g., \citealt{crifo}) and a trigonometric
parallax of 68.2$\pm$1.8~mas \citep{gj}, which corresponds to a distance of
14.7$\pm$0.4~pc. The trigonometric distance of LHS 1604 is consistent with
its spectroscopic distance of 12.8$\pm$1.8~pc \citep{crifo}.
With $J=11.2$ \citep{pb03} and the trigonometric distance, 
we derived $M_{\rm J} = 10.36$.
Based on the BT-Settl models, 
the detection of lithium in LHS~1604, its color $I-J = 2.55$ \citep{pb03}
and its $J$-band absolute magnitude imply that the source has
an age range of 100$-$150~Myr and a substellar mass (see Fig.~\ref{f5}).
Using this age range and the $J$-band absolute magnitude of the source, 
we derived its mass to be 55$-$66~$M_{\rm J}$. 
We then adopt an average mass of $\sim$61~$M_{\rm J}$ to the source.
Using the BANYAN II tool, the proper motion measurement in \citet{pb03}
and a radial velocity of $-$11.9$\pm$2.0~km~s$^{-1}$ \citep{deshpande},
we found that LHS 1604 has a 100\% probability candidate member of young field.
LHS~1604 is therefore a young field BD within 15~pc.

{\it DENIS0518$-$3101 (M6.5):} This low-proper motion 
dwarf was identified by \citet{pb03}. It has a spectral type of M6.5 \citep{crifo}.
\citet{mac12} detected a strong radio emission at 8.5~GHz 
from DENIS0518$-$3101. 
No trigonometric parallax and radial velocity of the source are available.
Using the BT-Settl models, 
its color $I-J=2.3$ and our marginal detection of lithium 
suggest its age to be $\leq$150~Myr.
This age limit places an upper limit of $\sim$73~$M_{\rm J}$ to the mass
of the object. 
DENIS0518$-$3101 is therefore a young BD candidate. At this point,
deeper observations are required to confirm the lithium presence in
the object and thus to confirm its substellar nature.  
Using the BAYAN II tool and the low-proper motion measurement in
\citet{pb03}, we found that the source has 
a 93.3\% membership probability of the Columba association
and a kinematic distance of 46$\pm$9~pc. 
If the source is at this distance, its absolute magnitude will be
$M_{\rm J} = 8.55$ ($J=11.87$). 
However, according to the BT-Settl models, 
this absolute magnitude implies that the source should have an age range of 10$-$20~Myr 
(see Fig.~\ref{f5}) that is not consistent with an age of 42$^{+6}_{-4}$~Myr
determined for Columba \citep{bell}.
Therefore, measurements of radial velocity and/or parallax of DENIS0518$-$3101
are clearly needed to
determine its membership in young moving groups.

{\it SIPS J1809$-$7613 (M5.0):} The source has 
an estimated spectral type of M5.0 and a spectroscopic
distance of 10.4~pc (Table~\ref{tab_SpT}).  
No measurements of trigonometric parallax and radial velocity 
of the object have been reported so far.
The clear detection of lithium in SIPS J1809$-$7613
will place the source in the lithium region (see Fig.\ref{f5}).
This therefore indicates that the real absolute magnitude of the source
must be brighter than its magnitude derived from the spectroscopic distance. 
According to the BT-Settl models,
the lithium presence in the object and its color $I-J=1.82$ (Table~\ref{tab_SpT})
indicate its age $\leq$~80~Myr. 
This thus places an upper limit of 95~$M_{\rm J}$ to the mass of SIPS J1809$-$7613.
Using the proper motion measurement from \citet{deacon},
the BAYAN II tool implies that the object has 
a 76.2\% membership probability of the $\beta$ Pic moving group
and a kinematic distance of 27$\pm$3~pc. 
Assuming the source at this distance, 
with $J=9.87$ (Table~\ref{tab_SpT}) we then derived $M_{\rm J} = 7.71$.
With such a magnitude,
the color-magnitude diagram (Fig.\ref{f5}) 
suggests that the age of the object should be in the range of 10$-$20~Myr.
This is likely consistent with the age of $\beta$ Pic of 24$\pm$3~Myr \citep{bell}.
Using the estimated age range and the DENIS color $I-J$ of the source,
we derived its mass to be 81$-$85~$M_{\rm J}$.  
We then adopt an average mass of 83~$M_{\rm J}$ for SIPS J1809$-$7613. 
The object is therefore a young VLM star.

{\it DENIS2022$-$5645 (M5.5):} This low-proper motion dwarf was discovered by \citet{pb03}
and spectroscopically classified as an M5.5 in \citet{crifo}. 
The source has no trigonometric parallax and radial velocity
measured so far.
The presence of lithium in DENIS2022$-$5645 indicates that the object
should locate in the lithium region as shown in Fig.\ref{f5}.
Therefore, the real absolute magnitude of the source actually
must be brighter than its magnitude derived from the spectroscopic distance. 
Based on the BT-Settl models,
our Li~I detection in the source and its DENIS color $I-J = 2.06$ \citep{pb03}
indicate that the source age should be $\leq$~120 Myr. 
This age constraint places an upper limit of $\sim$79~$M_{\rm J}$ to the mass of
DENIS2022$-$5645, which is very close to the substellar boundary. 
The source is therefore a young BD candidate.
Using the proper motion measurement in \citet{pb03} and the BAYAN II tool,
we found that the object has a 59.9\% probability candidate member of 
Tucana-Horologium and a kinematic distance of 53$\pm4$~pc.
Assuming the source at this kinematic distance,
we then derived $M_{\rm J} = 8.13$ ($J=11.75$). 
With such an absolute magnitude,
the BT-Settl models (Fig.\ref{f5}) indicate that
the source should have an age of $\leq$$\sim$10~Myr. 
This age estimate, however, disagrees with the age of 45$\pm$4~Myr determined for 
Tucana-Horologium \citep{bell}.
Therefore, radial velocity and/or parallax measurements
are additionally required for DENIS2022$-$5645 
to determine its membership in young moving groups and thus to confirm its substellar nature.

\begin{figure*}
   \centering
    \includegraphics[width=10.0cm,angle=-90]{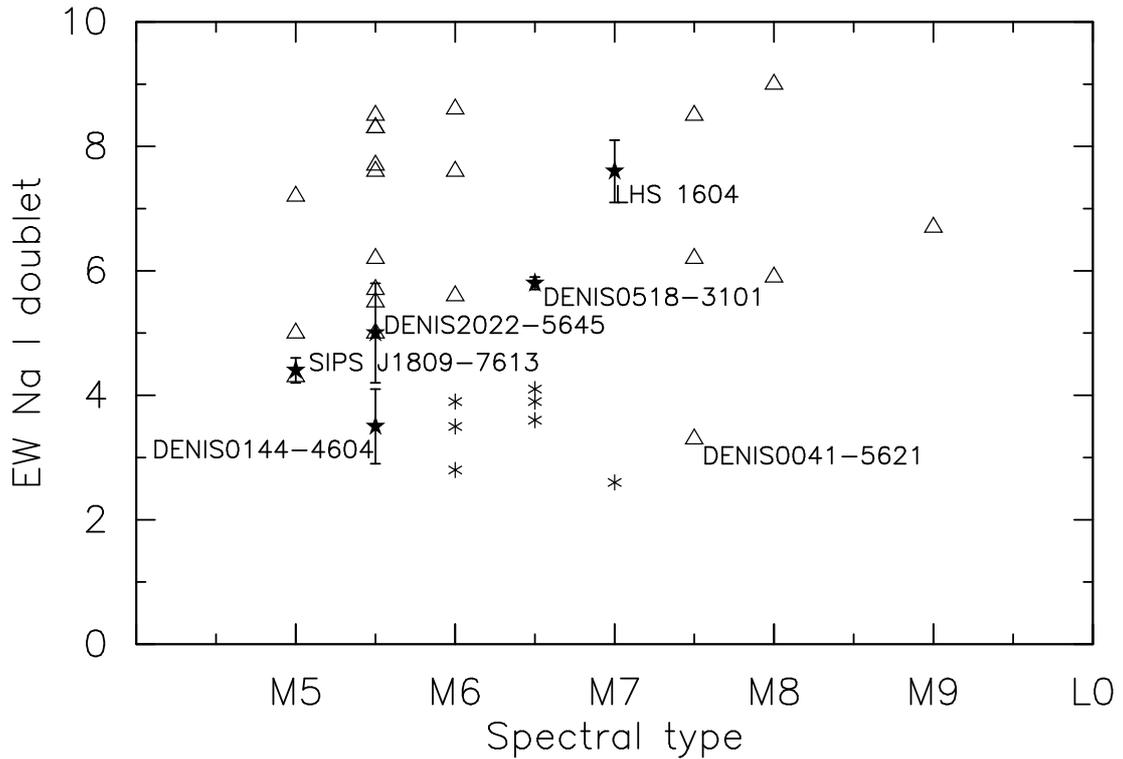}
    \caption{Na~I (8183/8199 \AA) equivalent width versus spectral type diagram for our 28 late-M dwarfs and
7 Upper Sco candidate members (asterisk symbols, \citealt{martin10}). Five late-M dwarfs (star symbols) 
with detected lithium in this paper 
are shown as well as DENIS0041$-$5621 with lithium detection reported in \citet{reiners09}
and \citet{bur15}.
}
\label{f4}
\end{figure*}

\begin{figure*}
   \centering
    \includegraphics[width=10.0cm,angle=-90]{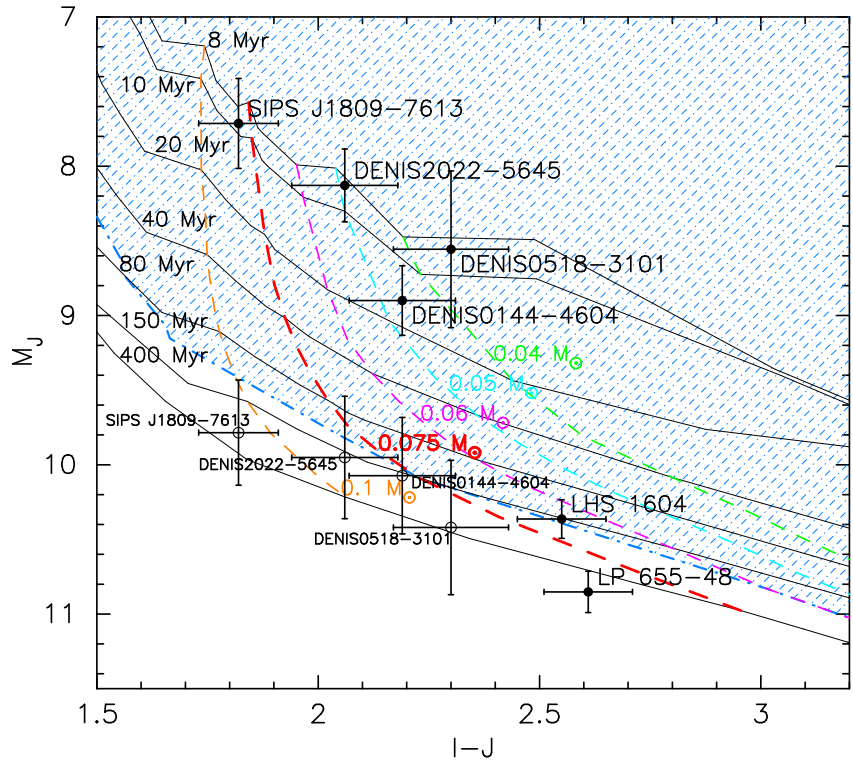}
    \caption{$J$-band absolute magnitude versus color $I-J$ diagram for the five 
late-M dwarfs with detected lithium. 
Isochrones and mass tracks from the CIFIST2011 BT-Settl atmosphere models \citep{allard13} 
are plotted. 
The lithium region is marked with the blue hatched area.
For the objects with no parallax measurements 
(DENIS0144$-$4604, DENIS0518$-$3101,
SIPS J1809$-$7613 and DENIS2022$-$5645),
the open and solid circles represent their absolute magnitudes derived
from spectroscopic and kinematic distances, respectively
(see Sect.~5.1 for further details). 
LP 655-48 is also shown (see Sect.~4.2 for discussion).}
\label{f5}
\end{figure*}
\subsection{A search for debris disks around three nearby VLM objects}
We also searched for debris disks around three late-M ($\geq$M7.0) and very nearby
sources (LHS~1604, M7.0), LP~655-48 (M7.5) and DENIS0517$-$3349 (M8.0). 
LHS~1604 is a young field BD as discussed in Sect.~5.1. 
Debris disks are made of planetesimals
left over the process of planet formation. In debris disks, dust is continuously
produced by collision and evaporation of planetesimals. Therefore,
the detection of dust emission
(i.e., debris disks) around relatively young dwarfs (ages $\geq$$\sim$10~Myr) 
implies the presence of larger bodies around the dwarfs \citep{wyatt}. 

We observed the three targets at 850 $\mu$m with the SCUBA-2 bolometer array 
\citep{holland} at the James Clerk Maxwell Telescope. 
The data were obtained on 2016-03-31, 2016-04-01 and 2016-07-29 
(UT time)
during which time zenith opacities at 225~GHz were in the range of 0.5$-$0.7. 
The primary FWHM beam is about 13$''$ 
at the observed wavelength \citep{dempsey}. 
The data were reduced using the Dynamic Iterative Map Maker (DIMM) 
in the SMURF package from the STAR-LINK software \citep{jenness} 
using the blank-field recipe designed specifically for faint emission \citep{chapin}. 
Both CRL618 and Uranus were observed as flux calibrators 
and were found to be within the nominal values quoted by \citet{dempsey}. 
We note that the uncertainty in the absolute flux calibration is $\sim$5\%. 
We applied an additional correction factor of 10\% 
to the data to compensate for the flux lost during 
match-filtering 
(as applied to the data when running the blank-field reduction, 
e.g., \citealt{chen}).

We did not detect any dust emission at 850~$\mu$m from the sources.
We measured the rms of 2.7 mJy, 5.0 mJy and 4.4 mJy for the continuum maps 
of LHS~1604, LP~655-48 and DENIS0517$-$3349, respectively.
Based on 3~rms flux densities,
we then used the formula for optically thin dust to estimate
the upper limits  
to the dust mass of debris disks for M dwarfs (see \citealt{lestrade}):
$S_{\lambda}$ = $M_{\rm d}$ $\times$ $B(\lambda,T_{\rm g})$ $\times$ $\kappa_{\rm 850\mu m}/d^{2}$,
where $T_{\rm g}$ is the dust temperature, d is the distance to the source,
$\kappa_{\rm 850\mu m}$ = 1.7~cm$^{2}$ g$^{-1}$ is the mass absorption coefficient of the dust 
at 850 $\mu$m \citep{dent,shirley}. 
The trigonometric distances were used for LHS~1604 (14.7 pc, \citealt{gj}) 
and LP~655-48 (9.5 pc, \citealt{shkol12}), and the spectroscopic
distance for DENIS0517$-$3349 (12.1 pc, see Table~\ref{tab_EW}).
Assuming a mean dust temperature of 15~K \citep{lestrade}, 
we then derived upper limits to the dust mass
to be $\sim$4.3, 3.3 and 4.8 lunar masses for 
LHS~1604, LP~655-48 and DENIS0517$-$3349, respectively.

\section{Conclusion}
In this paper, we report our detections of lithium in five late-M dwarfs.
The results indicate our lithium detection rate of about 14\%.
Using the theoretical models, we estimated their masses.
Our mass estimates indicate that
DENIS0144$-$4604 (M5.5) and LHS~1604 (M7.0) are young BDs,
DENIS0518$-$3101 (M6.5) and DENIS2022$-$5645 (M5.5) are young BD candidates,
and  SIPS J1809$-$7613 (M5.0) is a young VLM star.
LHS~1604 is a young field BD at only 15~pc with an age range of 100$-$150 Myr.
Measurements of radial velocity and trigonometric parallax are needed for
the four remaining sources to confirm their substellar nature as well as their
membership in young moving groups.

\begin{acknowledgements}
This research is funded by Vietnam National University HoChiMinh City (VNU-HCM) 
under grant number C2014-28-01. 
The research based on the JCMT data is funded by Vietnam National Foundation 
for Science and Technology Development (NAFOSTED) under grant number 103.99-2015.108. 
EM acknowledges funding from the Spanish Ministry of 
Economy and Competitiveness (MINECO) under grant AYA-2015-69350-C3-1. 
We would like to thank the referee for useful comments.
The James Clerk Maxwell Telescope is operated by the East Asian Observatory on 
behalf of The National Astronomical Observatory of Japan, 
Academia Sinica Institute of Astronomy and Astrophysics, 
the Korea Astronomy and Space Science Institute, 
the National Astronomical Observatories of China and 
the Chinese Academy of Sciences (Grant No. XDB09000000), 
with additional funding support from the Science and 
Technology Facilities Council of the United Kingdom and 
participating universities in the United Kingdom and Canada.
This research has made use of the VizieR catalogue access tool, CDS,
Strasbourg, France. The original description of the VizieR service was
published in A\&AS 143, 23.
This research has made use of the SIMBAD database,
operated at CDS, Strasbourg, France.
\end{acknowledgements}


\begin{thebibliography}{}

\bibitem[Allard et al.(2013)]{allard13}
  Allard, F., Homeier, D., Freytag, B., Schaffenberger, W., \& Rajpurohit, A. S. 2013, MmSAI, 24, 128

\bibitem[Basri et al.(1996)]{basri96}
  Basri, G., Marcy, G. W.,  \&  Graham, J. R. 1996, ApJ, 458, 600

\bibitem[Basri(2000)]{basri00}
  Basri, G. 2000, ARA\&A, 38, 485

\bibitem[Bell et al.(2015)]{bell}
  Bell, C. P. M., Mamajek, E. E., \& Naylor, T. 2015, MNRAS, 454, 593

\bibitem[Bessell(1999)]{bessell99} 
  Bessell, M. S. 1999, PASP, 111, 1426

\bibitem[Berger et al.(2008a)]{berger08a}
  Berger, E., Gizis, J. E., Giampapa, M. S., Rutledge, R. E., et al. 2008a, ApJ, 673, 1080 
 
\bibitem[Berger et al.(2008b)]{berger08b}
  Berger, E., Basri, G., Gizis, J. E., Giampapa, M. S., et al. 2008b, ApJ, 676, 1307

\bibitem[Biller et al.(2006)]{bill}
  Biller, B. A., Kasper, M., Close, L. M., Brandner, W., \& Kellner, S. 2006, ApJ, 641, L141

\bibitem[Bopp(1974)]{bopp}
  Bopp, B. W. 1974, PASP, 86, 281

\bibitem[Boyd et al.(2011)]{boyd}
  Boyd, M. R., Henry, T. J., Jao, W.-C., Subasavage, J. P., \& Hambly, N. C. 2011, AJ, 142, 92

\bibitem[Brandt \& Huang(2015)]{brandt}
  Brandt, T. D., \& Huang, C. X. 2015, \apj, 807, 58

\bibitem[Burgasser et al.(2015)]{bur15} 
  Burgasser, A. J., Logsdon, S. E., Gagn\'e, J., Bochanski, J. J., et al. 2015, ApJS, 220, 18

\bibitem[Chabrier et al.(1996)]{chab96}
  Chabrier, G., Baraffe, I., \& Plez, B. 1996, ApJ, 459, L91

\bibitem[Chabrier \& Baraffe(2000)]{chab}
  Chabrier, G., \& Baraffe, I. 2000, ARA\&A, 38, 337

\bibitem[Chapin et al.(2013)]{chapin}
  Chapin, E. L., Berry, D. S., Gibb, A. G., Jenness, T., Scott, D., Tilanus, R. P. J., Economou,    
  F., \& Holland, W. S. 2013, MNRAS, 430, 2545 

\bibitem[Chen et al.(2013)]{chen}  
  Chen, C. C., Cowie, L. L., Barger, A. J., Casey, C. M., Lee, N., Sanders, D. B., Wang, W. H.,
  \& Williams, J. P. 2013, ApJ, 776, 131

\bibitem[Crifo et al.(2005)]{crifo}
  Crifo, F., Phan-Bao, N., Delfosse, X., Forveille, T., Guibert, J., Mart\'{\i}n, E. L., 
  Reyl\'e, C. 2005, A\&A, 441, 653

\bibitem[Cruz \& Reid(2002)]{cruz02}
  Cruz, K. L., \& Reid, I. N. 2002, AJ, 123, 2828

\bibitem[Deacon \& Hambly(2007)]{deacon}
  Deacon, N. R., \& Hambly, N. C. 2007, A\&A, 468, 163

\bibitem[Dempsey et al.(2013)]{dempsey}
  Dempsey, J. T., Friberg, P., Jenness, T., Tilanus, R. P. J., et al. 2013, MNRAS, 430, 2534 

\bibitem[Dent et al.(2000)]{dent}
  Dent, W. R. F., Walker, H. J., Holland, W. S., \& Greaves, J. S. 2000, MNRAS, 314, 702 

\bibitem[Deshpande et al.(2012)]{deshpande}
  Deshpande, R., Mart\'{\i}n, E. L., Montgomery, M. M., et al. 2012, AJ, 144, 99

\bibitem[Faherty et al.(2012)]{faherty}
 Faherty, J. K., Burgasser, A. J., Walter, F. M., Van der Bliek, N., et al. 2012, ApJ, 752 

\bibitem[Favata et al.(1997)]{favata}
 Favata, F., Micela, G., \& Sciortino, S. 1997, A\&A, 322, 131

\bibitem[Filippazzo et al.(2015)]{filip}
 Filippazzo, J. C., Rice, E. L., Faherty, J., et al. 2015, ApJ, 810, 158

\bibitem[Finch et al.(2007)]{fin}
  Finch, C. T., Henry, T. J., Subasavage, J. P., Jao, W.-C., \& Hambly, N. C. 2007, AJ, 133, 2898

\bibitem[Gagn\'e et al.(2014)]{gagne14}
  Gagn\'e, J., Lafreni\`ere, D., Doyon, R., Malo, L., \& Artigau, E. 2014, ApJ, 783, 121

\bibitem[Gagn\'e et al.(2015)]{gagne15a}
  Gagn\'e, J., Lafreni\`ere, D., Doyon, R., Malo, L., \& Artigau, E. 2015, ApJ, 798, 73

\bibitem[Galvez-Ortiz et al.(2010)]{go10}
 G\'alvez-Ortiz, M. C., Clarke, J. R. A., Pinfield, D. J., Jenkins, J. S., et al. 2010, MNRAS, 409, 552

\bibitem[Gliese \& Jahrei\ss(1991)]{gj}
   Gliese, W., \& Jahrei\ss, H. 1991,  Preliminary Version of the
   Third Catalogue of Nearby Stars, as available at CDS Strasbourg

\bibitem[Hambly et al.(2004)]{hambly}
  Hambly, N. C., Henry, T. J., Subasavage, J. P., Brown, M. A., Jao, W.-C. 2004, AJ, 128, 437

\bibitem[Henry et al.(2004)]{henry}
  Henry, T. J., Subasavage, J. P., Brown, M. A., Beaulieu, T. D., Jao, W.-C., \& Hambly, N. C. 2004, AJ, 128, 2460

\bibitem[Holland al.(2013)]{holland}  
 Holland, W. S., Bintley, D., Chapin, E. L., Chrysostomou, A., et al. 2013, MNRAS, 430, 2513

\bibitem[Jenness et al.(2011)]{jenness}
 Jenness, T., Berry, D., Chapin, E., Economou, F., Gibb, A., \& Scott, D. 2011, ASPC, 442, 281

\bibitem[L\'epine(2008)]{lepine}
  L\'epine, S. 2008, AJ, 135, 2177

\bibitem[Lestrade et al.(2006)]{lestrade}
  Lestrade, J.-F., Wyatt, M. C., Bertoldi, F., Dent, W. R. F., \& Menten, K. M. 2006, 
  A\&A, 460, 733

\bibitem[Luyten(1979)]{luyten}
  Luyten, W. J. 1979, Catalogue of stars with proper motions exceeding
  0.$''$5 annually (LHS) (Minneapolis: University of Minnesota)

\bibitem[Lyo et al.(2004)]{lyo}
  Lyo, A-R., Lawson, W. A., \& Bessell, M. S. 2004, MNRAS, 355, 363
    
\bibitem[Magazz\`u et al.(1993)]{magazzu93} 
Magazz\`u A.,  Mart\'{\i}n, E. L., \& Rebolo, R. 1993, ApJ, 404, L17 

\bibitem[Malo et al.(2013)]{malo}
  Malo, L., Doyon, R., Lafreni\`ere, D., Artigau, E., Gagn\'e, J., Baron, F., \& Riedel, A. 2013,  
  ApJ, 762, 88

\bibitem[Mart\'{\i}n et al.(1994)]{martin94} 
  Mart\'{\i}n, E. L., Rebolo, R., \& Magazz\`u A. 1994, ApJ, 436, 262 

\bibitem[Mart\'{\i}n et al.(1996)]{martin96} 
  Mart\'{\i}n E. L., Rebolo, R., \& Zapatero Osorio, M. R., 1996, ApJ, 469, 706

\bibitem[Mart\'{\i}n et al.(1999a)]{martin99a} 
  Mart\'{\i}n E. L., Basri G., \& Zapatero Osorio M. R., 1999a, AJ, 118, 1005

\bibitem[Mart\'{\i}n et al.(1999)]{martin99b} 
  Mart\'{\i}n E. L., Delfosse X., Basri G., Goldman B., Forveille T., Zapatero Osorio M. R., 1999b, AJ, 118, 2466

\bibitem[Mart\'{\i}n et al.(2010)]{martin10} 
  Mart\'{\i}n E. L., Phan-Bao, N., Bessell, M., Delfosse, X., et al. 2010, A\&A, 517, 53 

\bibitem[McLean et al.(2012)]{mac12}
  McLean, M., Berger, E., \& Reiners, A. 2012, ApJ, 746, 23

\bibitem[Messina et al.(2016)]{messina} 
  Messina, S., Lanzafame, A. C., Feiden, G. A., Millward, M., et al. 2016,  arXiv:1607.06634

\bibitem[Montagnier et al.(2006)]{mon}
  Montagnier, G., S\'egransan, D., Beuzit, J.-L., et al. 2006, A\&A, 460, L19

\bibitem[Murphy \& Lawson(2015)]{murphy}
  Murphy, S. J., \& Lawson, W. A. 2015, MNRAS, 447, 1267

\bibitem[Pavlenko et al.(1995)]{pav}
  Pavlenko, Y. V., Rebolo, R., Mart{\'{\i}}n E. L., \& Garc{\'{\i}}a L{\'o}pez, R. J. 
  1995, A\&A, 303, 807

\bibitem[Phan-Bao \& Bessell(2006)]{pb06}
  Phan-Bao N., \& Bessell M. S. 2006, A\&A, 446, 515

 \bibitem[Phan-Bao et al.(2001)]{pb01}
  Phan-Bao N. et al., 2001, A\&A, 380, 590

\bibitem[Phan-Bao et al.(2003)]{pb03}
  Phan-Bao N. et al., 2003, A\&A, 401, 959

\bibitem[Phan-Bao et al.(2005)]{pb05}
  Phan-Bao, N., Mart{\'{\i}}n, E. L., Reyl\'e, C., Forveille, T., \& Lim, J. 2005, A\&A, 439, L19

\bibitem[Phan-Bao et al.(2006a)]{pb06a}
  Phan-Bao, N., Bessell, M. S., Mart{\'{\i}}n, E. L., et al. 2006a, MNRAS, 366, L40

\bibitem[Phan-Bao et al.(2006b)]{pb06b}
  Phan-Bao, N., Forveille, T., Mart{\'{\i}}n, E. L., \& Delfosse, X. 2006b, ApJ, 645, L153

\bibitem[Phan-Bao et al.(2008)]{pb08}
  Phan-Bao, N., Bessell, M. S., Mart{\'{\i}}n, E. L., et al. 2008, MNRAS, 383, 831 

\bibitem[Phan-Bao et al.(2009)]{pb09}
  Phan-Bao, N., Lim, J., Donati, J.-F., Johns-Krull, C. M., \& Mart{\'{\i}}n, E. L. 2009, ApJ, 704, 1721

\bibitem[Pokorny et al.(2003)]{pok}
  Pokorny, R. S., Jones, H. R. A., \& Hambly, N. C. 2003, A\&A, 397, 575

\bibitem[Rajpurohit et al.(2013)]{raj}
  Rajpurohit, A. S., Reyl\'e, C., Allard, F., Homeier, D., Schultheis, M., 
  Bessell, M. S., \& Robin, A. C. 2013, A\&A, 556, 15

\bibitem[Rebolo et al.(1992)]{rebolo92}
  Rebolo R., Mart{\'{\i}}n E. L., \& Magazz\`u A. 1992, ApJ, 389, L83

\bibitem[Rebolo et al.(1995)]{rebolo95}
  Rebolo R., Zapatero Osorio M. R., \& Mart{\'{\i}}n E. L. 1995, Nature, 377, 129

\bibitem[Rebolo et al.(1996)]{rebolo96}
 Rebolo R., Mart{\'{\i}}n E. L., Basri, G., Marcy, G. W. \&  Zapatero Osorio M. R. 1996, 
 ApJ, 469, L53

\bibitem[Reid et al.(2002)]{reid02} 
  Reid, I. N., Kirkpatrick, J. D., Liebert, J., Gizis, J. E., Dahn C. C., \&
  Monet, D. G. 2002, AJ, 124, 519

\bibitem[Reiners \& Basri(2009)]{reiners09} 
 Reiners, A., \& Basri, G. 2009, ApJ, 705, 1416

\bibitem[Reiners et al.(2010)]{reiners10} 
 Reiners, A., Seifahrt, A., \& Dreizler, S. 2010, A\&A, 513, L9

\bibitem[Schmidt et al.(2007)]{schmidt}
  Schmidt, S. J., Cruz, K. L., Bongiorno, B. J., Liebert, J., \& Reid, I. N. 2007, AJ, 133, 2258

\bibitem[Schlieder et al.(2012)]{schlieder}
  Schlieder, J. E., L\'epine, S., Rice, E., Simon, M., Fielding, D., \& 
  Tomasino, R. 2012, AJ, 143, 114

\bibitem[Shirley et al.(2011)]{shirley}
  Shirley, Y. L., Huard, T. L., Pontoppidan, K. M., Wilner, D. J., Stutz, A. M., 
  Bieging, J. H., \& Evans, N. J. II 2011, ApJ, 728, 143 

\bibitem[Shkolnik et al.(2009)]{shkol09}
 Shkolnik, E. L., Liu, M. C., \& Reid, I. N. 2009, ApJ, 699, 649

\bibitem[Shkolnik et al.(2012)]{shkol12}
 Shkolnik, E. L., Anglada-Escud\'e, G., Liu, M. C., Bowler, B. P., Weinberger, A. J., 
 Boss, A. P., Reid, I. N., \& Tamura, M. 2012, ApJ, 758, 56

\bibitem[Shortridge et al.(2004)]{shortridge}
 Shortridge, K., Meyerdierks, H., Currie, M., Clayton, C., et al. 2004, Starlink User Note 86

\bibitem[Song et al.(2002)]{song}
  Song, I., Bessell, M. S., \& Zuckerman, B. 2002, ApJ, 581, L43

\bibitem[Subasavage et al.(2005)]{suba}
  Subasavage, J. P., Henry, T. J., Hambly, N. C., Brown, M. A., \& Jao, W.-C. 2005, AJ, 129, 413

\bibitem[Thackrah et al.(1997)]{thack}
  Thackrah, A., Jones, H., \& Hawkins, M. 1997, MNRAS, 284, 507

\bibitem[Tinney(1998)]{tinney98}
  Tinney, C.G. 1998, MNRAS, 296, L42

\bibitem[Torres et al.(2008)]{torres08}
  Torres, C. A. O., Quast, G. R., Melo, C. H. F., \& Sterzik, M. F. 2008, 
  in Young Nearby Loose Associations, Vol. I ed. B. Reipurth (The Southern Sky ASP
  Monograph Publications; San Francisco, CA: ASP), 757

\bibitem[West et al.(2008)]{west08}
  West, A. A., Hawley, S. L., Bochanski, J. J., Covey, K. R., Reid, I. N., Dhital, S., 
  Hilton, E. J., \& Masuda, M. 2008, AJ, 135, 785

\bibitem[West et al.(2011)]{west}
  West, A. A., Morgan, D. P., Bochanski, J. J., Andersen, J. M., et al. 2011, AJ, 141, 97

\bibitem[Wroblewski \& Torres(1991)]{wro}
  Wroblewski, H., \& Torres, C. 1991, A\&AS, 91, 129

\bibitem[Wyatt(2008)]{wyatt}
  Wyatt, M. C. 2008, ARA\&A, 46, 339 

\end{thebibliography}
\end{document}